%% file: main.tex
\documentclass[12pt]{article}
\usepackage[dvips]{graphicx}

\usepackage{amssymb}
\usepackage{amsmath}
\usepackage{latexsym}

\newcommand{\zh}{$Z_{\mathrm{H}}$}
\newcommand{\ah}{$A_{\mathrm{H}}$}
\newcommand{\wh}{$W_{\mathrm{H}}$}
\newcommand{\whpm}{$W_{\mathrm{H}}^{\pm}$}
\newcommand{\wpm}{$W^{\pm}$}
\newcommand{\eh}{$e_{\mathrm{H}}$}
\newcommand{\zhah}{$A_{\mathrm{H}}Z_{\mathrm{H}}$}

\newcommand{\zhzh}{$Z_{\mathrm{H}}Z_{\mathrm{H}}$}
\newcommand{\zhahahahbb}{$A_{\mathrm{H}}Z_{\mathrm{H}} \to A_{\mathrm{H}} A_{\mathrm{H}} bb$}
\newcommand{\nnhnnbb}{$\nu \nu h \to \nu \nu bb$}
\newcommand{\zhnnbb}{$Zh \to \nu \nu bb$}
\newcommand{\zznnbb}{$ZZ \to \nu \nu bb$}
\newcommand{\nnznnbb}{$\nu \nu Z \to \nu \nu bb$}
\newcommand{\gzgbb}{$\gamma Z \to \gamma bb$}
\newcommand{\ww}{$W^{+}W^{-}$}
\newcommand{\nnww}{$\nu\bar{\nu}W^{+}W^{-}$}
\newcommand{\eeww}{$e^{+}e^{-} W^{+}W^{-}$}
\newcommand{\enwz}{$e \nu_{e} WZ$}
\newcommand{\zww}{$Z W^{+}W^{-}$}
\newcommand{\misspt}{$^{\mathrm{miss}}p_{\mathrm{T}}$}
\newcommand{\pt}{$p_{\mathrm{T}}$}

\newcommand{\wwqqqq}{$W^{+}W^{-} \to qqqq$}
\newcommand{\nnwwnnqqqq}{$\nu\bar{\nu}W^{+}W^{-} \to \nu\bar{\nu} qqqq$}
\newcommand{\eewweeqqqq}{$e^{+}e^{-} W^{+}W^{-} \to e^{+}e^{-} qqqq$}
\newcommand{\zwwnnqqqq}{$ZW^{+}W^{-} \to \nu \bar{\nu} qqqq$}
\newcommand{\enwzenqqqq}{$e \nu_{e} W Z \to e \nu_{e} qqqq$}

\begin{document}

\begin{titlepage}
  \begin{center}

    \hfill KEK Preprint 2008-43 \\
    \hfill UT-HET 018 \\
    \hfill ICRR-Report-534 \\

    \vspace{0.4cm}
    {\large\bf Precision Measurements of Little Higgs Parameters \\
               at the International Linear Collider}
    \vspace{0.8cm}

    {\bf Eri Asakawa}$^{(a),}$\footnote{E-mail: eri@post.kek.jp},
    {\bf Masaki Asano}$^{(b),}$\footnote{E-mail: masano@icrr.u-tokyo.ac.jp},
    {\bf Keisuke Fujii}$^{(c),}$\footnote{E-mail: keisuke.fujii@kek.jp},
    \\
    {\bf Tomonori Kusano}$^{(d),}$\footnote{E-mail: kusano@awa.tohoku.ac.jp},
    {\bf Shigeki Matsumoto}$^{(e),}$\footnote{E-mail: smatsu@sci.u-toyama.ac.jp},
    {\bf Rei Sasaki}$^{(d),}$\footnote{E-mail: rei@awa.tohoku.ac.jp},
    \\
    {\bf Yosuke Takubo}$^{(d),}$\footnote{E-mail: takubo@neutrino.kek.jp},
    and
    {\bf Hitoshi Yamamoto}$^{(d),}$\footnote{E-mail: yhitoshi@awa.tohoku.ac.jp}

    \vspace{0.8cm}

    {\it
     $^{(a)}${Institute of Physics, Meiji Gakuin University, Yokohama, Japan} \\
     $^{(b)}${Institute for Cosmic Ray Research (ICRR), University of Tokyo, Kashiwa, Japan} \\
     $^{(c)}${High Energy Accelerator Research Organization (KEK), Tsukuba, Japan} \\
     $^{(d)}${Department of Physics, Tohoku University, Sendai, Japan} \\
     $^{(e)}${Department of Physics, University of Toyama, Toyama, Japan} \\
    }

    \vspace{0.8cm}

    \abstract{We investigate a possibility of precision measurements for parameters of the Littlest Higgs model with T-parity at the International Linear Collider (ILC). The model predicts new gauge bosons (\ah, \zh, and \wh), among which the heavy photon (\ah) is a candidate for dark matter. The masses of these new gauge bosons strongly depend on the vacuum expectation value that breaks a global symmetry of the model. Through Monte Carlo simulations of the processes: $e^+ e^- \rightarrow A_{\mathrm{H}} Z_{\mathrm{H}}$ and $e^+ e^- \rightarrow W_{\mathrm{H}}^+ W_{\mathrm{H}}^-$, we show how precisely the masses can be determined at the ILC for a representative parameter point of the model. We also discuss the determination of the Little Higgs parameters and its impact on the future measurement of the thermal abundance of the dark matter relics in our universe.}

  \end{center}
\end{titlepage}

\setcounter{footnote}{0}


\input Introduction
\input Model
\input Setup
\input Analysis
\input Discussions
\input Summary

\vspace{1.0cm}
\hspace{0.2cm} {\bf Acknowledgments}
\vspace{0.5cm}

The authors would like to thank all the members of the ILC physics subgroup \cite{Ref:subgroup} for useful discussions. They are grateful to the Minami-tateya group for the help extended in the early stage of the event generator preparation. This work is supported in part by the Creative Scientific Research Grant (No. 18GS0202) of the Japan Society for Promotion of Science and the JSPS Core University Program.

\end{document}

%% file: Introduction.tex
\section{Introduction} 
\label{sec:intro}

There is no doubt that the Higgs boson is the most important particle not only for the confirmation of the Standard Model (SM) but also for the exploration of physics beyond the SM. Quadratically divergent corrections to the Higgs mass term suggest that new physics should appear at the scale around 1 TeV. However, electroweak precision measurements require that the scale is larger than ${\cal O}(10)$ TeV in order not to conflict with the measurements \cite{Barbieri}. This problem is called the little hierarchy problem, and many people expect that new physics involves some mechanism to solve the problem.

There are a number of scenarios for new physics involving such a mechanism. The most famous one is the supersymmetric scenario. Recently, alternative one called the Little Higgs scenario \cite{Arkani-Hamed:2001nc, Arkani-Hamed:2002qy} has been proposed. In this scenario, the Higgs boson is regarded as a pseudo Nambu-Goldstone boson associated with a global symmetry at some higher scale. Though the symmetry is not exact, its breaking is specially arranged to cancel quadratically divergent corrections to the Higgs mass term at 1-loop level. This is called the Little Higgs mechanism. As a result, the scale of new physics can be as high as 10 TeV without a fine-tuning on the Higgs mass term.  Due to the symmetry, the scenario necessitates the introduction of new particles such as heavy gauge bosons and top partners.

It is also known that most of Little Higgs models still suffer from severe constraints from electroweak precision measurements due to direct couplings among a new heavy gauge boson and SM particles \cite{difficulty}. In order to resolve the problem, a $Z_2$ symmetry called T-parity is imposed on the models \cite{Cheng:2003ju}-\cite{Low:2004xc}. Under the parity, new particles are assigned to be T-odd (i.e. with a T-parity of $-1$), while the SM particles are T-even. Thanks to the symmetry, dangerous interactions mentioned above are prohibited. Furthermore, the lightest T-odd particle is stable and provides a good candidate for dark matter. In this article, we focus on the Littlest Higgs model with T-parity as a simple and typical example of models implementing both the Little Higgs mechanism and T-parity\footnote{We assume that T-parity is an exact symmetry. It depends on the UV completions whether the T-parity is an exact symmetry or not \cite{Hill:2007nz, Krohn:2008ye}.} \cite{Cheng:2004yc}-\cite{Hubisz:2004ft}. Heavy photon plays the role of dark matter in this model \cite{Hubisz:2004ft, Asano:2006nr}.

In order to test the Little Higgs model, precise determinations of properties of Little Higgs partners are mandatory, because these particles are directly related to the cancellation of quadratically divergent corrections to the Higgs mass term. In particular, measurements of heavy gauge boson masses are quite important. Since heavy gauge bosons acquire mass terms through the breaking of the global symmetry mentioned above, precise measurements of their masses allow us to determine the most important parameter of the model, namely the vacuum expectation value of the breaking. Furthermore, because the heavy photon is a candidate for dark matter, the determination of its property gives a great impact not only on particle physics but also on astrophysics and cosmology. At the Large Hadron Collider (LHC), top partners are expected to be copiously produced, and their properties will be determined accurately \cite{{Meade:2006dw},Matsumoto:2008fq}. However, it is difficult to determine the properties of heavy gauge bosons at the LHC, because they have no color charge \cite{Cao:2007pv}. 

On the other hand, the International Linear Collider (ILC) will provide an ideal environment to measure the properties of heavy gauge bosons. The ILC is the future electron-positron linear collider for the next generation of the high energy frontier physics. At the ILC, electrons and positrons are accelerated by two opposing linear accelerators installed in an about 30 km long underground tunnel, and are brought into collision with a center of mass energy of 500 GeV-1 TeV. Heavy gauge bosons are expected to be produced in a clean environment at the ILC, which enables us to determine their properties precisely. In this article, we study the sensitivity of the measurements to the Little Higgs parameters at the ILC based on a realistic Monte Carlo simulation. In addition, from the simulation results, we estimate the capability of the ILC to determine the thermal abundance of the dark matter (heavy photon) relics in our universe.

This article is organized as follows. In the next section, we briefly review the Littlest Higgs model with T-parity. Simulation framework such as a representative point in the parameter space of the model and the simulation tools used in our study are presented in section \ref{sec:mctool}. Details of analysis for heavy gauge boson productions at the ILC are discussed in section \ref{sec:analysis}, where we show expected measurement accuracies of the heavy gauge boson properties at both $\sqrt{s}$ = 500 GeV and 1 TeV. In section \ref{sec:discussion}, we will show how powerful the ILC is to determine the Little Higgs parameters based on the simulation results. Connection to cosmology from the ILC experiment is also discussed in this section. Section \ref{sec:summary} is devoted to summary.

%% file: Model.tex
\section{Littlest Higgs model with T-parity}
\label{sec:Model}

In this section, we briefly review the Littlest Higgs model with T-parity, in particular focusing on gauge-Higgs and lepton sectors of the model. (For general reviews of Little Higgs models, see Refs. \cite{littlest_review, littlest}.) 

\subsection{Gauge-Higgs sector}

The Littlest Higgs model with T-parity is based on a non-linear sigma model describing an SU(5)/SO(5) symmetry breaking. The non-linear sigma field $\Sigma$ is
\begin{eqnarray}
 \Sigma = e^{2 i \Pi / f} \Sigma_0,
\end{eqnarray}
where $f \sim {\cal O}(1)$ TeV is the vacuum expectation value of the breaking. The Nambu-Goldstone (NG) boson matrix $\Pi$ and the direction of the breaking $\Sigma_0$ are
\begin{eqnarray}
 \Pi
 =
 \begin{pmatrix}
                   0 & H  /\sqrt{2} & \Phi         \\
  H^\dagger/\sqrt{2} &            0 & H^T/\sqrt{2} \\
        \Phi^\dagger & H^*/\sqrt{2} & 0            \\
 \end{pmatrix},
 \qquad
 \Sigma_0
 =
 \begin{pmatrix}
        0 & 0 & {\bf 1} \\
        0 & 1 &       0 \\
  {\bf 1} & 0 &       0 \\
 \end{pmatrix}.
 \label{pNG matrix}
\end{eqnarray}
Here, we omit the would-be NG fields in the $\Pi$ matrix. An [SU(2)$\times$U(1)]$^2$ subgroup in the SU(5) global symmetry is gauged, which is broken down to the diagonal subgroup identified with the SM gauge group SU(2)$_L\times$U(1)$_Y$. Due to the presence of the gauge interactions and Yukawa interactions introduced in the next subsection, the SU(5) global symmetry is not exact, and particles in the $\Pi$ matrix become pseudo NG bosons. Fourteen (= 24 $-$ 10) NG bosons are decomposed into representations ${\bf 1}_0 \oplus {\bf 3}_0 \oplus {\bf 2}_{\pm 1/2} \oplus {\bf 3}_{\pm 1}$ under the electroweak gauge group. The first two representations are real, and become longitudinal components of heavy gauge bosons when the [SU(2)$\times$U(1)]$^2$ is broken down to the SM gauge group. The other scalars ${\bf 2}_{\pm 1/2}$ and ${\bf 3}_{\pm 1}$ are a complex doublet identified with the SM Higgs field ($H$ in Eq.\ (\ref{pNG matrix})) and a complex triplet Higgs field ($\Phi$ in Eq.\ (\ref{pNG matrix})), respectively.

The kinetic term of the $\Sigma$ field is given as
\begin{eqnarray}
  {\cal L}_{\Sigma}
  =
  \frac{f^2}{8}{\rm Tr}
  \left|
    \partial_\mu \Sigma
    -
    i\sqrt{2}
    \left\{
      g  ({\bf W} \Sigma + \Sigma {\bf W}^T)
      +
      g' ({\bf B} \Sigma + \Sigma {\bf B}^T)
    \right\}
  \right|^2,
  \label{Kinetic}
\end{eqnarray}
where ${\bf W} = W^a_j Q_j^a$ (${\bf B} = B_j Y_j$) is the ${\rm SU(2)}_j$ (${\rm U(1)}_j$) gauge field and $g$ ($g'$) is the ${\rm SU(2)}_L$ (${\rm U(1)}_Y$) gauge coupling constant. With the Pauli matrix $\sigma^a$, the generator $Q_j$ and the hyper-charge $Y_j$ are given as
\begin{eqnarray}
 &&
 Q_1^a
 =
 \frac{1}{2}
 \left(
  \begin{array}{ccc}
   \sigma^a & 0 & 0
   \\
   0 & 0 & 0
   \\
   0 & 0 & 0
  \end{array}
 \right)~,
 \qquad
 Q_2^a
 =
 -\frac{1}{2}
 \left(
  \begin{array}{ccc}
   0 & 0 & 0
   \\
   0 & 0 & 0
   \\
   0 & 0 & \sigma^{a*}
  \end{array}
 \right)~,
 \nonumber \\
 \nonumber \\
 &&
 Y_1
 =
 {\rm diag}(3,3,-2,-2,-2)/10~,
 \qquad
 Y_2
 =
 {\rm diag}(2,2,2,-3,-3)/10~.
\end{eqnarray}
It turns out that the Lagrangian in Eq.\ (\ref{Kinetic}) is invariant under T-parity,
\begin{eqnarray}
  \Pi \leftrightarrow -\Omega \Pi \Omega,
  \qquad
  W^a_1 \leftrightarrow W^a_2,
  \qquad
  B_1 \leftrightarrow B_2;
  \qquad
  \Omega = {\rm diag}(1,1,-1,1,1).
\end{eqnarray}

This model contains four kinds of gauge fields. The linear combinations $W^a = (W^a_1 + W^a_2)/\sqrt{2}$ and $B = (B_1 + B_2)/\sqrt{2}$ correspond to the SM gauge bosons for the SU(2)$_L$ and U(1)$_Y$ symmetries. The other linear combinations $W^a_{\mathrm{H}} = (W^a_1 - W^a_2)/\sqrt{2}$ and $B_{\mathrm{H}} = (B_1 - B_2)/\sqrt{2}$ are additional gauge bosons called heavy gauge bosons, which acquire masses of ${\cal O}(f)$ through the SU(5)/SO(5) symmetry breaking. After the electroweak symmetry breaking with $\langle H \rangle = (0, v/\sqrt{2})^T$, the neutral components of $W^a_{\mathrm{H}}$ and $B_{\mathrm{H}}$ are mixed with each other and form mass eigenstates \ah~and \zh,
\begin{eqnarray}
  \begin{pmatrix}
    Z_{\mathrm{H}} \\ A_{\mathrm{H}}
  \end{pmatrix}
  =
  \begin{pmatrix}
    \cos \theta_{\mathrm{H}} & -\sin \theta_{\mathrm{H}} \\
    \sin \theta_{\mathrm{H}} &  \cos \theta_{\mathrm{H}}     
  \end{pmatrix}    
  \begin{pmatrix}
    W_{\mathrm{H}}^3 \\ B_{\mathrm{H}} 
  \end{pmatrix}.
\end{eqnarray}
The mixing angle $\theta_{\mathrm{H}}$ is given as
\begin{eqnarray}
  \tan \theta_{\mathrm{H}}
  =
  - \frac{2m_{12}}
  {m_{11} - m_{22} + \sqrt{(m_{11} - m_{22})^2 + 4m_{12}^2}}
  \sim
  - 0.15\frac{v^2}{f^2},
\end{eqnarray}
where $m_{11} = g^2 f^2 (c_f^2 + 7)/8$, $m_{12} = g g^{\prime } f^2 (1 - c_f^2)/8$, $m_{22} = g^{\prime 2} f^2 (5c_f^2 + 3)/40$, and $c_f = \cos (\sqrt{2}v/f)$. Since the mixing angle is considerably suppressed, \ah~is dominantly composed of $B_{\mathrm{H}}$. Masses of gauge bosons are given by
\begin{eqnarray}
  m_W^2 &=& \frac{g^2}{4} f^2 (1 - c_f)
  \simeq \frac{g^2}{4} v^2,
  \\
  m_Z^2 &=& \frac{g^2 + g^{\prime 2}}{4} f^2 (1 - c_f)
  \simeq \frac{g^2 + g^{\prime 2}}{4} v^2,
  \\
  m_{W_{\mathrm{H}}}^2 &=& \frac{g^2}{4} f^2 (c_f + 3)
  \simeq g^2 f^2,
  \label{eq:mwh} \\
  m_{Z_{\mathrm{H}}}^2
  &=&
  \frac{1}{2}
  \left(m_{11} + m_{22} + \sqrt{(m_{11} - m_{22})^2 + 4m_{12}^2}\right)
  \simeq g^2 f^2,
  \label{eq:mzh} \\
  m_{A_{\mathrm{H}}}^2
  &=&
  \frac{1}{2}
  \left(m_{11} + m_{22} - \sqrt{(m_{11} - m_{22})^2 + 4m_{12}^2}\right)
  \simeq 0.2 g^{\prime 2} f^2.
  \label{m(A_H)}
\end{eqnarray}
As expected from the definitions of \ah, \zh, and \wh, the new heavy gauge bosons behave as T-odd particles, while SM gauge bosons are T-even.

Scalar potential terms for $H$ and $\Phi$ fields are radiatively generated \cite{Arkani-Hamed:2002qy,Hubisz:2004ft},
\begin{eqnarray}
 V(H, \Phi)
 =
 \lambda f^2{\rm Tr}\left[\Phi^\dagger\Phi\right]
 -
 \mu^2H^\dagger H
 +
 \frac{\lambda}{4}\left(H^\dagger H\right)^2
 +
 \cdots.
 \label{Potential}
\end{eqnarray}
Main contributions to $\mu^2$ come from logarithmically divergent corrections at 1-loop level and quadratically divergent corrections at 2-loop level. As a result, $\mu^2$ is expected to be smaller than $f^2$. The triplet Higgs mass term, on the other hand, receives quadratically divergent corrections at 1-loop level, and therefore is proportional to $f^2$. The quartic coupling $\lambda$ is determined by the 1-loop effective potential from gauge and top sectors. Since both $\mu$ and $\lambda$ depend on parameters at the cutoff scale $\Lambda \simeq 4\pi f$, we treat them as free parameters. The mass of the triplet Higgs boson $\Phi$ is given by $m_\Phi^2 = \lambda f^2 = 2m_h^2f^2/v^2$, where $m_h$ is the mass of the SM Higgs boson. The triplet Higgs boson is T-odd, while the SM Higgs is T-even.

The gauge-Higgs sector of the model is composed of the kinetic term of the $\Sigma$ field in Eq.\ (\ref{Kinetic}) and the potential terms in Eq.\ (\ref{Potential}) in addition to appropriate kinetic terms of gauge fields $W^a_j$, $B_j$ and gluon $G$. It can be seen that the heavy photon \ah~is considerably lighter than other T-odd particles. Since the stability of \ah~is guaranteed by the conservation of T-parity, it becomes a good candidate for dark matter.

\subsection{Lepton sector}

\begin{table}[t]
  \center{
    \begin{tabular}{|c|c||c|c||c|c|}
      \hline
      $l^{(1)}$ & $({\bf 2}, -3/10; {\bf 1}, -1/5)$ &
      $l^{(2)}$ & $({\bf 1}, -1/5; {\bf 2}, -3/10)$ &
      $e_R$ & $({\bf 1}, -1/2 ; {\bf 1}, -1/2 )$ \\
   \hline
  \end{tabular}
  }
  \caption{\small Quantum number of $[SU(2)\times U(1)]^2$ for particles in the lepton sector.}
  \label{table:charges}
\end{table}

To implement T-parity, two SU(2) doublets $l^{(1)}$ and $l^{(2)}$ and one singlet $e_R$ are introduced for each SM lepton. The quantum numbers of these particles under the [SU(2)$\times$ U(1)]$^2$ gauge symmetry are shown in Table \ref{table:charges}. With these particles, Yukawa interactions invariant under gauge symmetries and T-parity turn out to be
\begin{eqnarray}
 {\cal L}_l^{\rm (Y)}
 &=& 
 i\frac{y_e}{4} f \epsilon_{ij} \epsilon_{xyz}
 \left[
   (\bar{\cal E}^{(2)})_x \Sigma_{i y} \Sigma_{j z} X
   -
   (\bar{\cal E}^{(1)} \Sigma_0)_x \tilde{\Sigma}_{i y} \tilde{\Sigma}_{j z}
   {\tilde{X}}
 \right]e_R
 \label{Yukawa_l1}
 \\
 &&
 -
 \kappa_l f
 \left(
  \bar{\cal N}^{(2)} \xi \Psi^c
  +
  \bar{\cal N}^{(1)} \Sigma_0 \Omega \xi^\dagger \Omega \Psi^c
 \right)
 +
 h.c.~,
 \label{Yukawa_l2}
\end{eqnarray}
where ${\cal N}^{(n)}$ are incomplete $SU(5)$ multiplets, ${\cal N}^{(1)} = (l^{(1)}, 0, 0)^T$, ${\cal N}^{(2)} = (0, 0, l^{(2)})^T$, ${\cal E}^{(n)} = -\sigma^2 {\cal N}^{(n)}$, and $l^{(n)} = -\sigma^2 (\nu_L^{(n)}, e_L^{(n)})^T$, while $\Psi_l^c$ is a complete multiplet of $SO(5)$, $\Psi_l^c = (\tilde{l}^c, \chi^c_l, l^c)^T$.  The indices $x,y,z$ run from 3 to 5 whereas $i,j = 1, 2$. For $X$, there are two possible choices: $X = (\Sigma_{33})^{-1/4}$ and $X = (\Sigma^\dagger_{33})^{1/4}$ \cite{Chen:2006cs}.  With $\tilde{\Sigma} = \Sigma_0 \Omega \Sigma^{\dagger} \Omega \Sigma_0$ and $\Sigma \equiv \xi^2 \Sigma_0$, these fields transform under T-parity as
\begin{eqnarray}
 {\cal N}^{(1)} \leftrightarrow - \Sigma_0 {\cal N}^{(2)},
\quad
 \Psi_l^c \leftrightarrow -\Psi_l^c ,
\quad
 \Sigma 
 \leftrightarrow 
 \tilde{\Sigma},
\quad
 X \leftrightarrow \tilde{X},
\quad
 \xi \leftrightarrow \Omega \xi^{\dagger} \Omega.
\label{Tparity_l}
\end{eqnarray}

The linear combination $l_{SM} = (l^{(1)} - l^{(2)})/\sqrt{2}$ gives the left-handed SM lepton, which acquires the Dirac mass term with $e_R$ in Eq.\ (\ref{Yukawa_l1}) through the electroweak symmetry breaking. On the other hand, another linear combination $l_{\mathrm{H}} = (l^{(1)} + l^{(2)})/\sqrt{2}$ acquires the Dirac mass term of ${\cal O}(f)$ with $l^c = -\sigma^2 (\nu^{c (n)}, e^{c (n)})^T$ in Eq.\ (\ref{Yukawa_l2}). As expected in Eq.\ (\ref{Tparity_l}), the heavy lepton $l_{\mathrm{H}}$ behaves as a T-odd particle, while the SM lepton $l_{SM}$ is T-even. The masses of the heavy leptons depend on $\kappa_l$,
\begin{eqnarray}
  m_{e_{\mathrm{H}}} &=& \sqrt{2} \kappa_l f, 
\qquad
  m_{\nu_{\mathrm{H}}} = \left( \frac{\sqrt{2}+\sqrt{1+c_f}}{2} \right)
              \kappa_l f
         \simeq \sqrt{2} \kappa_l f.
\end{eqnarray}

The lepton sector of the model is composed of the Yukawa interactions above and appropriate kinetic terms of above leptons involving gauge interactions associated with gauge charges shown in Table \ref{table:charges}.

%% file: Setup.tex
\section{Simulation framework}
\label{sec:mctool}

\subsection{Representative point in the parameter space}

In order to perform a numerical simulation at the linear collider, we need to choose a representative point in the parameter space of the Littlest Higgs model with T-parity. Firstly, the model parameters should satisfy the current electroweak precision data. In addition, the cosmological observation of dark matter relics also gives important information. Thus, we consider not only the electroweak precision measurements but also the WMAP observation \cite{Komatsu:2008hk} to choose a point in the parameter space.

We have calculated the $\chi^2$-function for observables:
\begin{eqnarray}
 \chi^2
 =
 \sum_i
 \frac{\left({\cal O}_{\rm obs}^{(i)} - {\cal O}_{\rm th}^{(i)}\right)^2}
      {\left(\Delta {\cal O}_{\rm obs}^{(i)}\right)^2},
 \label{chi2}
\end{eqnarray}
where ${\cal O}_{\rm obs}^{(i)}$, ${\cal O}_{\rm th}^{(i)}$, and $\Delta {\cal O}_{\rm obs}^{(i)}$ are an observed value, its theoretical prediction, and the error of the observation for observable ${\cal O}^{(i)}$. For the observed values, the following four observables are considered: the $W$ boson mass ($m_W = 80.412 \pm 0.042$ GeV), the weak mixing angle ($\sin^2\theta^{\rm lept}_{\rm eff} = 0.23153 \pm 0.00016$), leptonic width of the $Z$ boson ($\Gamma_l = 83.985 \pm 0.086$ MeV) \cite{LEPSLC}, and the relic abundance of dark matter ($\Omega_{\rm DM} h^2 = 0.119 \pm 0.009$) \cite{de Austri:2006pe}. On the other hand, theoretical predictions for these observables depend on three  model parameters; $f$, $\lambda_2$\footnote{Unlike the masses of heavy gauge bosons, those of top partners depend on not only $f$ but also $\lambda_2$. Since the top partners are irrelevant to our analysis, we 
will not discuss the parameter. See Ref.\cite{Hubisz:2004ft} for more details.}, $m_h$. (For the detailed expressions for the predictions, see \cite{Hubisz:2005tx, Asano:2006nr}). For theoretical predictions, the fine structure constant at the $Z$ pole ($\alpha^{-1}(m_Z) = 128.950$), the top quark mass ($m_t = 172.7$ GeV) \cite{Arguin:2005cc}, the $Z$ boson mass ($m_Z = 91.1876$ GeV), and the Fermi constant ($G_F = 1.16637 \times 10^{-5}$ GeV$^{-2}$) \cite{Yao:2006px} have been used as input parameters. For the $f$ parameter, the region $f < 570$ GeV, which corresponds to $m_{A_{\mathrm{H}}} < m_W$, is unattractive because the pair annihilation of \ah~into a gauge-boson pair is kinematically forbidden. 

Using the $\chi^2$ function, we have selected a representative point $(f, m_h, \lambda_2)$ $=$ (580 GeV, 134 GeV, 1.5). At the representative point, we have obtained $\Omega_{\rm DM} h^2$ of 1.05. Notice that no fine-tuning is needed at the sample point to keep the Higgs mass on the electroweak scale 
\cite{Hubisz:2005tx,Matsumoto:2008fq}. The masses of the heavy gauge bosons and the triplet Higgs boson at the representative point are summarized in Table \ref{table:point}. It can be seen that all the heavy gauge bosons are lighter than 500 GeV, which allows us to consider their pair production at the ILC. 

\begin{table}[t]
  \center{
    \begin{tabular}{|c|c|c|c|}
      \hline
      $f$ (GeV) & $m_h$ (GeV) & $\lambda_2$ & $\kappa_l$ \\
      \hline
      580 & 134 & 1.5 & 0.5 \\
      \hline 
      \hline
      $m_{A_{\mathrm{H}}}$ (GeV) & $m_{W_{\mathrm{H}}}$ (GeV) &
      $m_{Z_{\mathrm{H}}}$ (GeV) & $m_{\Phi}$ (GeV) \\
      \hline
      81.9 & 368 & 369 & 440 \\
      \hline
    \end{tabular}
  }
  \caption{\small Representative point used in our simulation study.}
  \label{table:point}
\end{table}

Here, we add a comment on the parameter $\kappa_l$ in Eq.\ (\ref{Yukawa_l2}), because cross sections to produce the heavy gauge bosons depend on the masses of the heavy leptons as well as the other parameters mentioned above. Though the parameter $\kappa_l$ is not directly related to the observables used in the $\chi^2$-analysis, it is also constrained by collider experiments. Since small $\kappa_l$ means the existence of light \eh, too small $\kappa_l$ 
has been ruled out by non-observation of new charged particles. On the other hand, large $\kappa_l$ is disfavored because it gives a large contribution to four-Fermi operators \cite{Hubisz:2005tx,Cao:2007pv}. Therefore, $\kappa_l$ is expected to be ${\cal O}(1)$, and we set $\kappa_l = 0.5$ in this article\footnote{When $\kappa_l < 0.45$, heavy leptons $e_{\mathrm{H}}$ and $\nu_{\mathrm{H}}$ are lighter than heavy gauge bosons \wh~and \zh. Collider signals will be changed drastically in that case~\cite{Cao:2007pv}.}.

\begin{table}[t]
  \center{
    \begin{tabular}{|c||c|c|c|c|}
      \hline
      $\sqrt{s}$ &
      $e^+e^- \rightarrow A_{\mathrm{H}}Z_{\mathrm{H}}$ &
      $e^+e^- \rightarrow Z_{\mathrm{H}}Z_{\mathrm{H}}$ &
      $e^+e^- \rightarrow W_{\mathrm{H}}^+W_{\mathrm{H}}^-$ \\
      \hline
      500 GeV & 1.91 (fb) & --- & --- \\
      \hline
      1 TeV & 7.42 (fb) & 110 (fb) & 277 (fb) \\
      \hline
    \end{tabular}
  }
  \caption{\small Cross sections for the production of heavy gauge bosons.}
  \label{table:Xsections}
\end{table}

\begin{figure}[t]
 \begin{center}
  \scalebox{0.75}{\includegraphics*{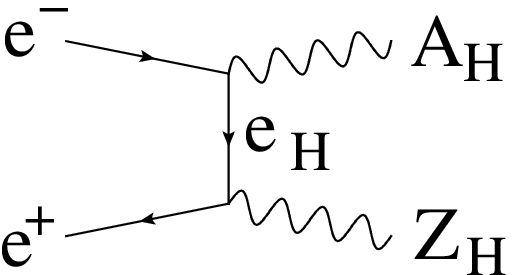}}
  \qquad\qquad
  \scalebox{0.75}{\includegraphics*{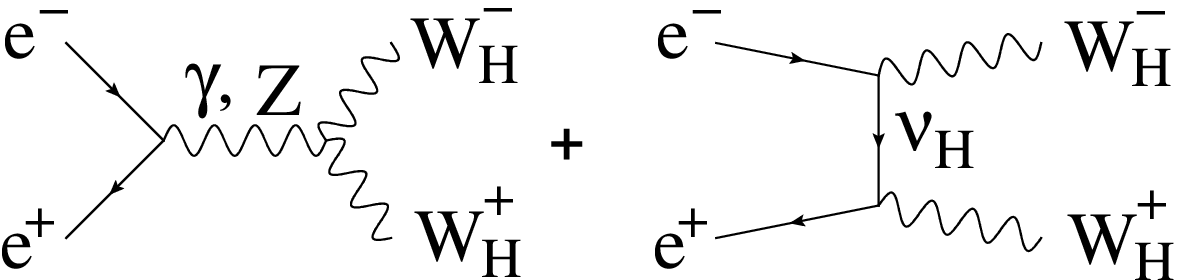}}
  \caption{\small Diagrams for signal processes; $e^+e^- \rightarrow A_{\mathrm{H}}Z_{\mathrm{H}}$ and $e^+e^- \rightarrow W_{\mathrm{H}}^+W_{\mathrm{H}}^-$.}
  \label{fig:signal diagrams}
 \end{center}
\end{figure}

There are four processes whose final states consist of two heavy gauge bosons: $e^+e^- \rightarrow$ $A_{\mathrm{H}}A_{\mathrm{H}}$, $A_{\mathrm{H}}Z_{\mathrm{H}}$, $Z_{\mathrm{H}}Z_{\mathrm{H}}$, and $W_{\mathrm{H}}^+ W_{\mathrm{H}}^-$. The first process is undetectable, thus not considered in this article\footnote{Furthermore, even if we consider the process $e^+e^- \rightarrow A_{\mathrm{H}}A_{\mathrm{H}}\gamma$, its cross section is strongly suppressed due to the small coupling between \ah~and leptons.}. The cross sections of the other processes are shown in Table \ref{table:Xsections}. It can be seen that the largest cross section is expected for the fourth process, which is open at $\sqrt{s} > 1$ TeV. On the other hand, because $m_{A_{\mathrm{H}}} + m_{Z_{\mathrm{H}}}$ is less than 500 GeV, the second process is important already at the $\sqrt{s} = 500$ GeV. We, hence, concentrate on $e^+e^- \rightarrow A_{\mathrm{H}}Z_{\mathrm{H}}$ at $\sqrt{s} = 500$ GeV and $e^+e^- \rightarrow W_{\mathrm{H}}^+W_{\mathrm{H}}^-$ at $\sqrt{s} = 1$ TeV. Feynman diagrams for the signal processes are shown in Fig. \ref{fig:signal diagrams}. Note that \zh~decays into $A_{\mathrm{H}} h$, and \whpm~decays into $A_{\mathrm{H}}W^\pm$ with almost 100\% branching fractions.

\subsection{Simulation tools}

We have used MadGraph \cite{madgraph} to generate signal events of the Little Higgs model, while Standard Model events have been generated by Physsim \cite{physsim}\footnote{Initial-state radiation and beamstrahlung have not been included in the event generations.}. We have ignored the finite crossing angle between the electron and positron beams. In both event generations, helicity amplitudes were calculated using the HELAS library \cite{helas}, which allows us to deal with the effect of gauge boson polarizations properly. Phase space integration and the generation of parton 4-momenta have been performed by BASES/SPRING \cite{bases}. Parton showering and hadronization have been carried out by using PYTHIA6.4 \cite{pythia}, where final-state tau leptons are decayed by TAUOLA \cite{tauola} in order to handle their polarizations correctly.

\begin{table}
 \center{
  \begin{tabular}{lcr}
   \hline
   Detector & Performance & Coverage \\
   \hline
   Vertex detector &
   $\delta_{b} \leq 5 \oplus 10/ p \beta \sin^{3/2}\theta$ ($\mu$m) &
   $|\cos\theta| \leq 0.93$
   \\
   Central drift chamber &
   $\delta p_{t}/p_{t}^{2} \leq 5 \times 10^{-5}$ (GeV/c)$^{-1}$ &
   $|\cos\theta| \leq 0.98$
   \\
   EM calorimeter &
   $\sigma_{E}/E = 17\% / \sqrt{E} \oplus 1\%$ &
   $|\cos\theta| \leq 0.99$
   \\
   Hadron calorimeter &
   $\sigma_{E}/E = 45\% / \sqrt{E} \oplus 2\%$ &
   $|\cos\theta| \leq 0.99$
   \\
   \hline
  \end{tabular}
 }
 \caption{\small Detector parameters used in our simulation study.}
 \label{tb:GLD}
\end{table}

The generated Monte Carlo events have been passed to a detector simulator called JSFQuickSimulator, which implements the GLD geometry and other detector-performance related parameters \cite{glddod}. In the detector simulator, hits by charged particles at the vertex detector and track parameters at the central tracker are smeared according to their position resolutions, taking into account correlations due to off-diagonal elements in the error matrix. Since calorimeter signals are simulated in individual segments, a realistic simulation of cluster overlapping is possible. Track-cluster matching is performed for the hit clusters in the calorimeter in order to achieve the best energy flow measurements. The resultant detector performance in our simulation study is summarized in Table \ref{tb:GLD}.

%% file: Analysis.tex
\section{Results from simulation study} \label{sec:analysis}

In this section, we present some results from our simulation study for heavy gauge boson productions. The simulation has been performed at $\sqrt{s} =$ 500 GeV for the \zhah~production and at $\sqrt{s} =$ 1 TeV for the $W_{\mathrm{H}}^+ W_{\mathrm{H}}^-$ production with an integrated luminosity of 500 fb$^{-1}$ each.

\subsection{The \zhah~production}

\begin{figure}
 \begin{center}
  \includegraphics[width=10cm]{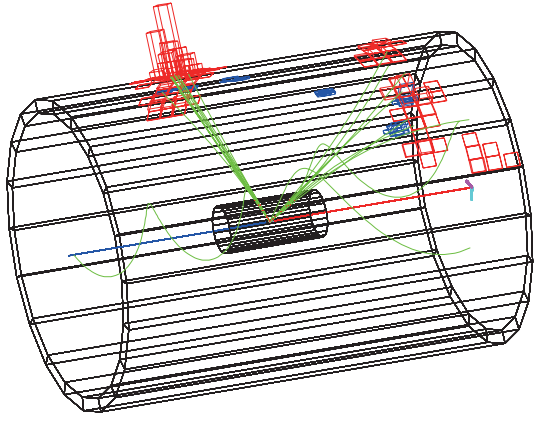}
 \end{center}
 \caption{\small A typical event of \zhah~in the simulator.}
 \label{fig:evtdsp}
\end{figure}

The heavy gauge bosons \ah~and \zh~are produced with the cross section of 1.9 fb at the center of mass energy of 500 GeV. Since \zh~decays into \ah~and the Higgs boson, the signature is a single Higgs boson in the final state, mainly 2 jets from $h \to b\bar{b}$ (with a 55\% branching ratio). We, therefore, define \zhahahahbb~as our signal event. For background events, contribution from light quarks was not taken into account because such events can be rejected to negligible level after requiring the existence of two $b$-jets, assuming a $b$-tagging efficiency of 80\% for $b$-jets with 15\% probability to misidentify a $c$-jet as a $b$-jet. This $b$-tagging performance was estimated by the full simulation assuming a typical ILC detector. Signal and background processes considered in this analysis are summarized in Table \ref{tb:zhevlst}. Figure \ref{fig:evtdsp} shows a typical \zhah~event as seen in the detector simulator.

\begin{table}
 \center{
  \begin{tabular}{l|r|r|r}
   \hline
   Process  & Cross sec. [fb] & \# of events & \# of events after all cuts \\
   \hline
   \zhahahahbb& 1.05            & 525          & 272             \\
   \nnhnnbb & 34.0            & 17,000       & 3,359           \\
   \zhnnbb  & 5.57            & 2,785        & 1,406           \\
   $tt \to WWbb$     & 496             & 248,000      & 264             \\
   \zznnbb  & 25.5            & 12,750       & 178           \\
   \nnznnbb & 44.3            & 22,150       & 167             \\
   \gzgbb   & 1,200           & 600,000      & 45               \\
   \hline
  \end{tabular}
 }
 \caption{\small Signal and backgrounds processes considered in the \zhah~analysis.}
 \label{tb:zhevlst}
\end{table}

The clusters in the calorimeters are combined to form a jet if the two clusters satisfy $y_{ij} < y_{\mathrm{cut}}$. $y_{ij}$ is defined as
\begin{equation}
 y_{ij} = \frac{2 E_{i} E_{j} (1 - \cos \theta_{ij})}{E_{\mathrm{vis}}^{2}},
\end{equation}
where $\theta_{ij}$ is the angle between momenta of two clusters, $E_{i(j)}$ are their energies, and $E_{\mathrm{vis}}$ is the total visible energy. All events are forced to have two jets by adjusting $y_{\mathrm{cut}}$. We have selected events with the reconstructed Higgs mass in a window of 100-140 GeV. In order to suppress the \nnhnnbb~background, the transverse momentum of the reconstructed Higgs bosons (\pt) is required to be above 80 GeV. This is because the Higgs bosons coming from the $WW$ fusion process, which dominates the \nnhnnbb~background, have \pt~mostly below W mass. Finally, multiplying the efficiency of double $b$-tagging ($0.8 \times 0.8 = 0.64$), we are left with 272 signal and 5,419 background events as shown in Table \ref{tb:zhevlst}, which corresponds to a signal significance of 3.7 ($= 272/\sqrt{5419}$) standard deviations. The indication of the new physics signal can hence be obtained at $\sqrt{s} = 500$ GeV.

The \ah~and \zh~boson masses can be estimated from the edges of the distribution of the reconstructed Higgs boson energies. This is because the maximum and minimum Higgs boson energies ($E_{\mathrm{max}}$ and $E_{\mathrm{min}}$) are written in terms of these masses,
\begin{eqnarray}
 E_{\mathrm{max}}
 &=& 
 \gamma_{Z_{\mathrm{H}}} E^{\ast}_{h}
 + 
 \beta_{Z_{\mathrm{H}}} \gamma_{Z_{\mathrm{H}}} p^{\ast}_{h},
 \nonumber \\ 
 E_{\mathrm{min}}
 &=& 
 \gamma_{Z_{\mathrm{H}}} E^{\ast}_{h}
 - 
 \beta_{Z_{\mathrm{H}}} \gamma_{Z_{\mathrm{H}}} p^{\ast}_{h},  
 \label{eq:eedge}
\end{eqnarray}
where $\beta_{Z_{\mathrm{H}}} (\gamma_{Z_{\mathrm{H}}})$ is the $\beta (\gamma)$ factor of the \zh~boson in the laboratory frame, while $E^{\ast}_{h} (p_{h}^{\ast})$ is the energy (momentum) of the Higgs boson in the rest frame of the \zh~boson. Note that $E^{\ast}_{h}$ is given as $(M_{Z_{\mathrm{H}}}^2 + M_h^2 - M_{A_{\mathrm{H}}}^2)/(2M_{Z_{\mathrm{H}}})$.

\begin{figure}
 \begin{center}
  \includegraphics[width=14cm]{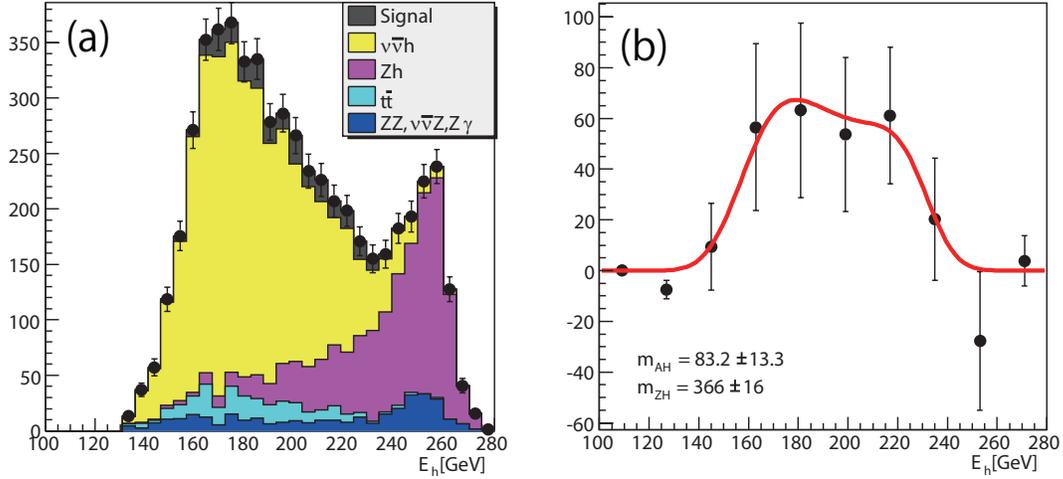}
 \end{center}
 \caption{\small (a) Energy distribution of the reconstructed Higgs bosons with remaining backgrounds after the mass cut. (b) Energy distribution of the Higgs bosons after subtracting the backgrounds. The distribution is fitted by a line shape function determined with a high statistics signal sample.}
 \label{fig:hene}
\end{figure}

The energy distribution of the reconstructed Higgs bosons with remaining backgrounds is depicted in Fig.\ref{fig:hene}(a). The signal distribution after backgrounds have been subtracted is shown in Fig.\ref{fig:hene}(b). The endpoints, $E_{\mathrm{max}}$ and $E_{\mathrm{min}}$, have been estimated by fitting the distribution with a line shape determined by a high statistics signal sample. The fit resulted in $m_{A_{\mathrm{H}}}$ and $m_{Z_{\mathrm{H}}}$ being $83.2 \pm 13.3$ GeV and $366.0 \pm 16.0$ GeV, respectively, which should be compared to their true values: 81.85 GeV and 368.2 GeV. 

\subsection{The $W_{\mathrm{H}}^+ W_{\mathrm{H}}^-$ production}

$W_{\mathrm{H}}^+ W_{\mathrm{H}}^-$ production has large cross section (277 fb) at the ILC with the center of mass energy of 1 TeV. Since \whpm~decays into \ah~and \wpm~ with the 100\% branching ratio, analysis procedure depends on the $W$ decay modes. In this analysis, we have used 4-jet final states from hadronic decays of two $W$ bosons, $W_{\mathrm{H}}^+W_{\mathrm{H}}^- \rightarrow A_{\mathrm{H}} A_{\mathrm{H}} qqqq$. Signal and background processes considered in the analysis are summarized in Table \ref{tb:whevlst}.

\begin{table}
 \center{
  \begin{tabular}{l|r|r|r}
   \hline
   Process                             & cross sec. [fb] & \# of events & \# of events after all cuts \\
   \hline
   $W_{\mathrm{H}}^+W_{\mathrm{H}}^- \rightarrow A_{\mathrm{H}}A_{\mathrm{H}}qqqq$ & 120             & 60,000       & 41,190          \\
   \wwqqqq                             & 1307            & 653,500      & 678             \\
   \eewweeqqqq                         & 490             & 245,000      & 46              \\
   \enwzenqqqq                         & 24.5            & 12,250       & 3,797           \\
   $Z_{\mathrm{H}}Z_{\mathrm{H}} \rightarrow A_{\mathrm{H}}A_{\mathrm{H}}qqqq$     & 18.8            & 9,400        & 213             \\
   \nnwwnnqqqq                         & 7.23            & 3,615        & 1,597           \\
   \zwwnnqqqq                          & 5.61            & 2,805        & 1,533           \\
   \hline
  \end{tabular}
  }
  \caption{\small Signal and background processes considered in the $W_{\mathrm{H}}^+ W_{\mathrm{H}}^-$ analysis.}
 \label{tb:whevlst}
\end{table}

All events have been reconstructed as 4-jet events by adjusting the cut on y-values. In order to identify the two $W$ bosons from \whpm~decays, two jet-pairs have been selected so as to minimize a $\chi^2$ function,
\begin{equation}
\chi^2
= 
(^{\mathrm{rec}}\mathrm{M}_{W1} -~^{\mathrm{tr}}\mathrm{M}_{W})^{2}/\sigma_{\mathrm{M}_{W}}^{2} 
+ 
(^{\mathrm{rec}}\mathrm{M}_{W2} -~^{\mathrm{tr}}\mathrm{M}_{W})^{2}/\sigma_{\mathrm{M}_{W}}^{2},
\end{equation}
where $^{\mathrm{rec}}\mathrm{M}_{W1(2)}$ is the invariant mass of the first (second) 2-jet system paired as a $W$ candidate, $^{\mathrm{tr}}\mathrm{M}_{W}$ is the true $W$ mass (80.4 GeV), and $\sigma_{\mathrm{M}_{W}}$ is the resolution for the $W$ mass (4 GeV). We required $\chi^2 < 26$ to obtain well-reconstructed events. 
Since \ah~bosons escape from detection resulting in a missing momentum, the missing transverse momentum (\misspt) of the signal peaks at around 175 GeV. We have thus selected events with \misspt~above 84 GeV. The numbers of events after the selection cuts are shown in Table \ref{tb:whevlst}. Notice that the \zhzh~and \eeww~events are reduced to a negligible level after imposing all the cuts. The number of remaining \ww, \enwz, \nnww, and \zww~background events is much smaller than that of the signal.

\begin{figure}
 \begin{center}
  \includegraphics[width=14cm]{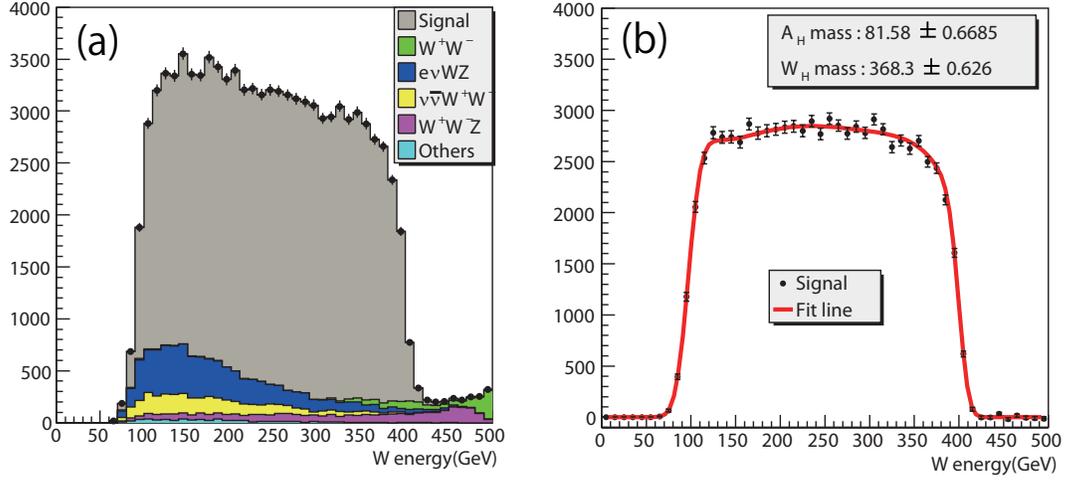}
 \end{center}
 \caption{\small (a) The energy distribution of the reconstructed $W$ bosons with remaining backgrounds after the selection cuts. (b) The energy distribution of the $W$ bosons after the subtraction of the backgrounds. The distribution is fitted by a line shape function determined with a high statistics signal sample.}
 \label{fig:wene}
\end{figure}

\begin{figure}
 \begin{center}
  \includegraphics[width=14cm]{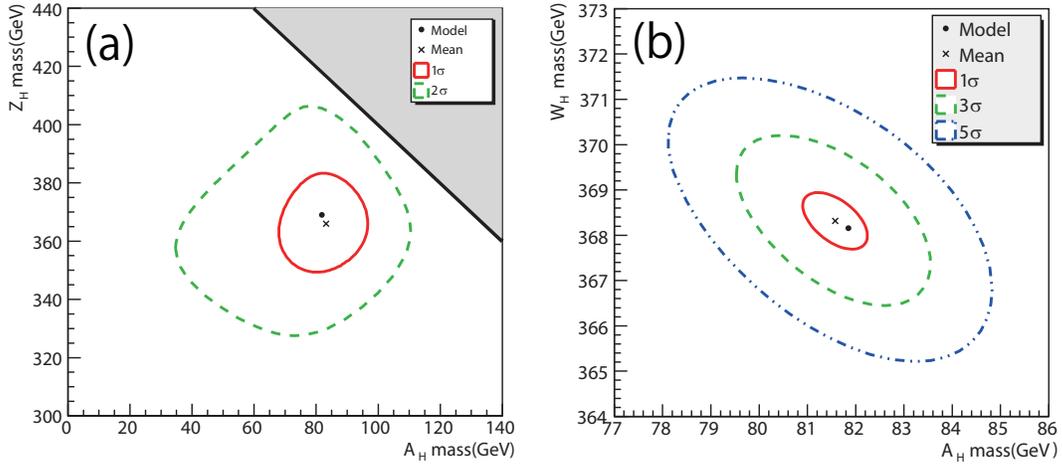}
 \end{center}
 \caption{\small Probability contours corresponding to (a) 1- and 2-$\sigma$ deviations from the best fit point in the \ah~and \zh~mass plane, and (b) 1-, 3-, and 5-$\sigma$ deviations in the \ah~and \wh~mass plane. The shaded area in (a) shows the unphysical region of $m_{A_{\mathrm{H}}} + m_{Z_{\mathrm{H}}} > 500$ GeV.}
 \label{fig:cntr}
\end{figure}

As in the case of the \zhah~production, the masses of \ah~ and \wh~bosons can be determined from the edges of the $W$ energy distribution. The energy distribution of the reconstructed $W$ bosons is depicted in Fig.\ref{fig:wene}(a). After subtracting the backgrounds from Fig.\ref{fig:wene}(a), the distribution has been fitted with a line shape determined by a high statistics signal sample as shown in Fig.\ref{fig:wene}(b). The fitted masses of \ah~and \wh~ bosons are $81.58 \pm 0.67$ GeV and $368.3 \pm 0.6$ GeV, respectively, which are to be compared to their input values: 81.85 GeV and 368.2 GeV. Figure \ref{fig:cntr} shows the probability contours for the masses of \ah~and \wh~at $1$ TeV together with that of \ah~and \zh~at 500 GeV. The mass resolution improves dramatically at $\sqrt{s} = 1 $ TeV, compared to that at $\sqrt{s} = 500$ GeV

\begin{figure}
 \begin{center}
  \includegraphics[width=8cm]{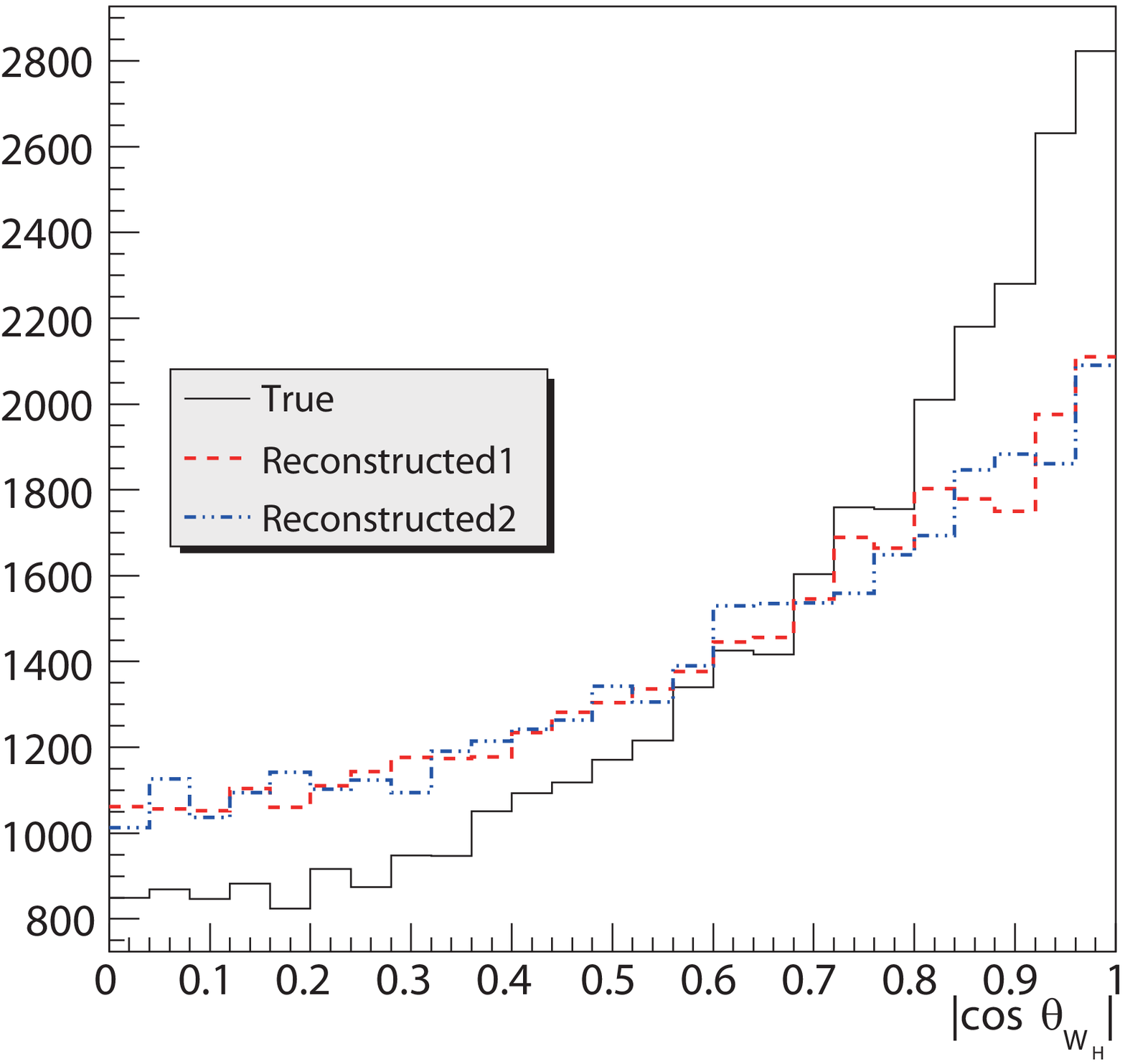}
 \end{center}
 \caption{\small Distribution of the production angles of \wh~bosons calculated from the momenta of the reconstructed $W$ bosons. Although there are two solutions, they have similar distributions.}
 \label{fig:wh_angle}
\end{figure}

The production angle of \wh~bosons can be calculated with 2-fold ambiguity from the momenta of $W$ bosons, assuming back-to-back production of $W_{\mathrm{H}}^+$ and $W_{\mathrm{H}}^-$. It turned out that the wrong solutions have a similar distribution to that of the correct ones. In Fig.\ref{fig:wh_angle}, we histogram the two solutions for the production angle, whose distribution is consistent with \wh~being spin-1 particle. 

\begin{figure}
 \begin{center}
  \includegraphics[width=8cm]{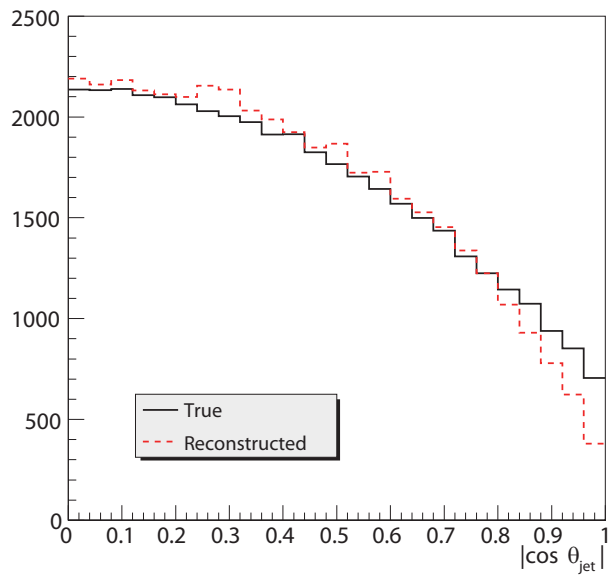}
 \end{center}
 \caption{\small Angular distribution of the reconstructed jets from $W$ in the $W$ helicity-frame.}
\label{fig:jet_angle}
\end{figure}

The angular distribution of jets in the helicity-frame of the parent \wpm~carries information on the polarization of the \wpm, from which we can extract information on the decay vertex of the parent particle. Figure \ref{fig:jet_angle} shows the angular distribution of the reconstructed jets in the helicity-frame of the reconstructed \wpm~bosons. The distribution indicates the dominance of the longitudinal \wpm~bosons, implying that this coupling arises from the electroweak symmetry breaking.

\begin{figure}
 \begin{center}
  \includegraphics[width=8cm]{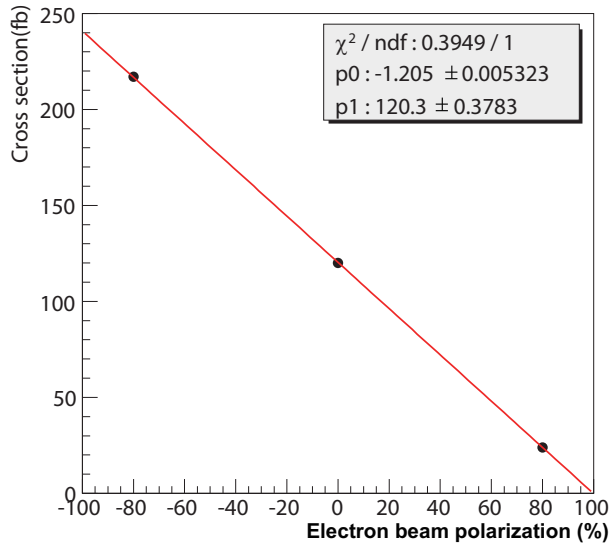}
 \end{center}
 \caption{\small The simulated cross section measurements for $e^+e^-\rightarrow W_{\mathrm{H}}^+W_{\mathrm{H}}^-$ with expected statistical errors, which are too small to be seen, as a function of the electron-beam polarization, where the positron-beam polarization was set to 0\%. The minus (plus) sign is for the left(right)-handed polarization.}
 \label{fig:xsec-pol}
\end{figure}

At the ILC, the electron beam, and optionally the positron beam as well, can be polarized. Changing the beam polarization, we can, hence, determine the $SU(2)_{\mathrm{L}}$ and $U(1)_{\mathrm{Y}}$ charges of \wh~through the measurements of the $e^+e^-\rightarrow W_{\mathrm{H}}^+W_{\mathrm{H}}^-$ cross sections. We have studied the measurement accuracy of the cross sections for the electron-beam polarizations of -80\%, 0\%, and +80\%, with the positron-beam polarization set to 0\%, where the minus (plus) sign is for the left(right)-handed polarization. Figure \ref{fig:xsec-pol} shows the simulated cross section measurements (data points with error bars which are too small to be seen) as a function of the electron-beam polarization together with a line representing the prediction by theory. Notice that the measured cross sections extrapolate to zero for the 100\% right-handed electron beam, indicating that \wh~ has $SU(2)_{\mathrm{L}}$ charge but no $U(1)_{\mathrm{Y}}$ charge. 

%% file: Discussions.tex
\section{Discussions}
\label{sec:discussion}

As shown in the previous section, the masses of the heavy gauge bosons \ah, \zh, and \wh~ can be determined very accurately at the ILC experiment. It is important to notice that these masses are obtained in a model-independent way, so that it is possible to test the Little Higgs model by comparing them with the theoretical predictions. Furthermore, since the masses of the heavy gauge bosons are from the vacuum expectation value $f$ as shown in Eq.\ (\ref{eq:mwh}), (\ref{eq:mzh}), and (\ref{m(A_H)}), it is also possible to accurately determine $f$, which is the most important parameter of the model. The parameter $f$ is determined to be $f = 576.0 \pm 25.0$ GeV from the process $e^+e^- \rightarrow A_{\mathrm{H}} Z_{\mathrm{H}}$ at $\sqrt{s} =$ 500 GeV, while $f = 580.0 \pm 0.7$ GeV from the process $e^+e^- \rightarrow W_{\mathrm{H}}^+ W_{\mathrm{H}}^-$ at $\sqrt{s} =$ 1 TeV. Note that the input value of $f$ is 580 GeV in our simulation study.

\begin{figure}
 \begin{center}
  \includegraphics[width=14cm]{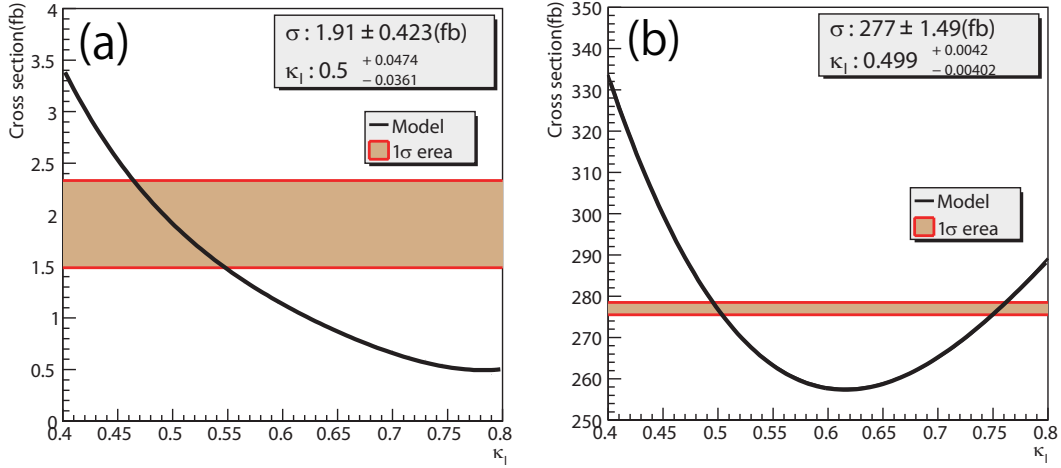}
 \end{center}
 \caption{\small The cross sections of (a) $e^+e^- \rightarrow A_{\mathrm{H}} Z_{\mathrm{H}}$ and (b) $e^+e^- \rightarrow W_{\mathrm{H}}^+ W_{\mathrm{H}}^-$ processes as a function of $\kappa_l$. One-sigma regions for the measurement accuracies of these cross sections are shown as shaded areas.}
 \label{fig:kappa}
\end{figure}

Another Little Higgs parameter $\kappa_l$ can also be determined from the results obtained in the previous section, because production cross sections for the heavy gauge bosons depend on the masses of heavy leptons $e_{\mathrm{H}}$ and $\nu_{\mathrm{H}}$. Figure \ref{fig:kappa} shows the cross-sections for $A_{\mathrm{H}}Z_{\mathrm{H}}$ production at $\sqrt{s} =$ 500 GeV and $W_{\mathrm{H}}^+W_{\mathrm{H}}^-$ production at $\sqrt{s} =$ 1 TeV as a function of $\kappa_l$. The measurement accuracies for these cross sections turned out to be 22.1\% at 500 GeV and 0.8\% at 1 TeV, which are shown as shaded regions in the figure. These cross section measurements constrain $\kappa_l$. Since the input value of $\kappa_l$ is 0.5 in our simulation study, these results correspond to the sensitivity to $\kappa_l$ of 9.5\% at 500 GeV and 0.8\% at 1 TeV. Although there are two possibilities for the value of $\kappa_l$ at 1 TeV, we can reject $\kappa_l$ of $\sim 0.75$ by the measurement at 500 GeV.

Once we obtain the Little Higgs parameters as above, it is possible to establish the connection between cosmology and the ILC experiment. Since the Little Higgs model has a candidate for WIMP dark matter \cite{Hubisz:2004ft, Asano:2006nr}, the most important physical quantity relevant to the connection is the thermal abundance of dark matter relics. It is well known that the abundance is determined by the annihilation cross section of dark matter \cite{Kolb:1990vq}. In the Little Higgs model, the cross section is determined by $f$ and $m_h$ in addition to well known gauge couplings \cite{Hubisz:2004ft}. The Higgs mass $m_h$ is expected to be measured very accurately at the ILC experiment 
\cite{GarciaAbia:1999kv}, so that it is quite important to measure $f$ accurately to predict the abundance.

\begin{figure}
 \begin{center}
  \includegraphics[width=10cm]{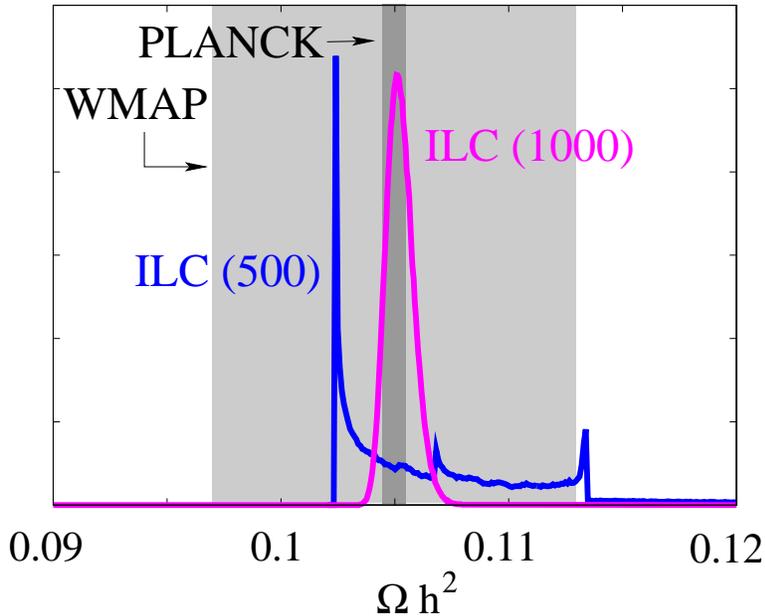}
 \end{center}
 \caption{\small The probability density of $\Omega h^2$ at $\sqrt{s} =$ 500 GeV and 1 TeV obtained from results in our simulation study. The measurement accuracies of cosmological observations (WMAP and PLANCK) are also shown as shaded regions.}
 \label{fig:Omegah2}
\end{figure}

Figure \ref{fig:Omegah2} shows how accurately the relic abundance can be determined at the ILC with the center of mass energies of 500 GeV and 1 TeV. The probability density of $\Omega h^2$, which is obtained from the results in the previous section, is depicted. As shown in the figure, the abundance will be determined with ${\cal O}$(10\%) accuracy even at $\sqrt{s} =$ 500 GeV, which is comparable to the WMAP observation. At $\sqrt{s} =$ 1 TeV, the accuracy will improve to 1\% level, which stands up to that expected for future cosmological observations such as from the PLANCK satellite \cite{Planck:2006uk}. The measurement accuracies of these cosmological observations are also shown in the figure in order to see the connection between the ILC experiment and cosmology.

Finally, we add comments on how our results compare with those expected from the LHC. Since the LHC is a hadron collider, it is not easy to identify heavy gauge boson productions \cite{Cao:2007pv}. However, new colored particles such as top partners will be copiously produced and their signals will be detected. When the masses of the top partners are around 1 TeV, the Little Higgs parameters such as $f$ will be determined with ${\cal O}(10)\%$ accuracy from the signals \cite{Matsumoto:2008fq}. It is then possible to determine the properties of the dark matter model-dependently, namely with the use of the relation between masses of the top partners and those of heavy gauge bosons. On the other hand, when the masses of the top partners are much heavier than 1 TeV, accurate determination of the Little Higgs parameters seems difficult at the LHC, though no realistic simulation in such a case is performed yet. While the LHC is exploring the colored sector of the Little Higgs model, the ILC will move on to cover the weak sector, thereby bringing us deeper understanding of the Little Higgs model.

%% file: Summary.tex
\section{Summary}
\label{sec:summary}

The Littlest Higgs Model with T-parity is one of the attractive candidates for physics beyond the Standard Model for it solves both the little hierarchy and dark matter problems simultaneously. One of the important predictions of the model is the existence of new heavy gauge bosons, where they acquire mass terms through the breaking of global symmetry necessarily imposed on the model. The determination of the masses are, hence, quite important to test the model. In this article, we have performed Monte Carlo simulations in order to estimate measurement accuracies of the masses (and of cross sections for heavy gauge boson productions) 
at the ILC for a representative parameter point of the model.

At the ILC with the center of mass energy of 500 GeV, it is possible to produce \ah~ and \zh~bosons with a signal significance of 3.7-sigma level. Furthermore, by observing the energy distribution of the Higgs bosons from the \zh~decays, the masses of these bosons can be determined with accuracies of 16.2\% for $m_{A_{\mathrm{H}}}$ and 4.3\% for $m_{Z_{\mathrm{H}}}$.

Once the ILC energy reaches $\sqrt{s}=$ 1 TeV, the process $e^+e^- \rightarrow W_{\mathrm{H}}^+W_{\mathrm{H}}^-$ opens. Since the cross section of the process is large, the masses of \wh~and \ah~can be determined as accurately as 0.8\% and 0.2\%, respectively. Using the process, it is also possible to confirm that the spin of \whpm~is consistent with one and the polarization of \wpm~from the \whpm~decay is dominantly longitudinal. Furthermore, we have shown that the gauge charges of the \wh~boson could be measured using a polarized electron beam.

We have also investigated how accurately the Little Higgs parameters can be determined at the ILC. From the results obtained in our simulation study, it turns out that the vacuum expectation value $f$ can be determined with accuracies of 4.3\% at $\sqrt{s}=$ 500 GeV and 0.1\% at 1 TeV. Another Little Higgs parameter $\kappa_l$, which is relevant to the lepton sector of the model, could also be estimated from production cross sections. At the ILC with 500 GeV and 1 TeV center of mass energies, $\kappa_l$ could be obtained within 9.5\% and 0.8\% accuracies, respectively.

Finally, we have discussed the connection between the ILC experiment and cosmology, focusing on the thermal abundance of dark matter relics, which is the most important physical quantity for the connection. We have found that the abundance can be determined with 10\% and 1\% levels at $\sqrt{s}=$ 500 GeV and 1 TeV, respectively. These accuracies are comparable to those of current and future cosmological observations for the cosmic microwave background, implying that the ILC experiment will play an essential role to understand the thermal history of our universe.

%% file: main.bbl
\begin{thebibliography}{9}

\bibitem{Barbieri}
  R.~Barbieri and A.~Strumia,
  Phys.\ Lett.\ B {\bf 433} (1998) 63;
  R.~Barbieri and A.~Strumia,
  arXiv:hep-ph/0007265.

\bibitem{Arkani-Hamed:2001nc}
  N.~Arkani-Hamed, A.~G.~Cohen and H.~Georgi,
  Phys.\ Lett.\ B {\bf 513} (2001) 232;
  N.~Arkani-Hamed, A.~G.~Cohen, E.~Katz, A.~E.~Nelson, T.~Gregoire and J.~G.~Wacker,
  JHEP {\bf 0208} (2002) 021.

\bibitem{Arkani-Hamed:2002qy}
  N.~Arkani-Hamed, A.~G.~Cohen, E.~Katz and A.~E.~Nelson,
  JHEP {\bf 0207} (2002) 034.

\bibitem{difficulty}
  C.~Csaki, J.~Hubisz, G.~D.~Kribs, P.~Meade and J.~Terning,
  Phys.\ Rev.\ D {\bf 67} (2003) 115002;
  J.~L.~Hewett, F.~J.~Petriello and T.~G.~Rizzo,
  JHEP {\bf 0310} (2003) 062;
  C.~Csaki, J.~Hubisz, G.~D.~Kribs, P.~Meade and J.~Terning,
  Phys.\ Rev.\ D {\bf 68} (2003) 035009;
  T.~Gregoire, D.~R.~Smith and J.~G.~Wacker,
  Phys.\ Rev.\ D {\bf 69} (2004) 115008;
  M.~C.~Chen and S.~Dawson,
  Phys.\ Rev.\ D {\bf 70} (2004) 015003;
  Z.~Han and W.~Skiba,
Phys.\ Rev.\ D {\bf 72} (2005) 035005;
  W.~Kilian and J.~Reuter,
  Phys.\ Rev.\ D {\bf 70} (2004) 015004.

\bibitem{Cheng:2003ju}
  H.~C.~Cheng and I.~Low,
  JHEP {\bf 0309} (2003) 051.

\bibitem{Cheng:2004yc}
  H.~C.~Cheng and I.~Low,
  JHEP {\bf 0408} (2004) 061.

\bibitem{Low:2004xc}
  I.~Low,
  JHEP {\bf 0410} (2004) 067.

\bibitem{Hubisz:2004ft}
  J.~Hubisz and P.~Meade,
  Phys.\ Rev.\ D {\bf 71} (2005) 035016,
  (For the correct paramter region consistent with the WMAP observation,
   see the figure in the revised vergion, hep-ph/0411264v3).


\bibitem{Asano:2006nr}
  M.~Asano, S.~Matsumoto, N.~Okada and Y.~Okada,
  Phys.\ Rev.\  D {\bf 75} (2007) 063506;
  A.~Birkedal, A.~Noble, M.~Perelstein and A.~Spray,
  Phys.\ Rev.\  D {\bf 74} (2006) 035002;
  M.~Perelstein and A.~Spray,
  Phys.\ Rev.\  D {\bf 75} (2007) 083519.
\bibitem{Hill:2007nz}
  C.~T.~Hill and R.~J.~Hill,
  Phys.\ Rev.\  D {\bf 75} (2007) 115009;
  C.~T.~Hill and R.~J.~Hill,
  Phys.\ Rev.\  D {\bf 76} (2007) 115014.

\bibitem{Krohn:2008ye}
  D.~Krohn and I.~Yavin,
  JHEP {\bf 0806} (2008) 092;
  C.~Csaki, J.~Heinonen, M.~Perelstein and C.~Spethmann, 
  arXiv:0804.0622 [hep-ph].
\bibitem{Meade:2006dw}
  P.~Meade and M.~Reece,
  Phys.\ Rev.\  D {\bf 74} (2006) 015010;
  C.~S.~Chen, K.~Cheung and T.~C.~Yuan,
  Phys.\ Lett.\  B {\bf 644} (2007) 158;
  A.~Freitas and D.~Wyler,
  JHEP {\bf 0611} (2006) 061;
  A.~Belyaev, C.~R.~Chen, K.~Tobe and C.~P.~Yuan,
  Phys.\ Rev.\  D {\bf 74} (2006) 115020;
  L.~Wang, W.~Wang, J.~M.~Yang and H.~Zhang,
  Phys.\ Rev.\  D {\bf 75} (2007) 074006;
  M.~S.~Carena, J.~Hubisz, M.~Perelstein and P.~Verdier,
  Phys.\ Rev.\  D {\bf 75} (2007) 091701;
  Q.~H.~Cao, C.~S.~Li and C.~P.~Yuan,
  Phys.\ Lett.\  B {\bf 668} (2008) 24;
  S.~Matsumoto, M.~M.~Nojiri and D.~Nomura,
  Phys.\ Rev.\  D {\bf 75} (2007) 055006;
  D.~Choudhury and D.~K.~Ghosh,
  JHEP {\bf 0708} (2007) 084;
  R.~Barcelo, M.~Masip and M.~Moreno-Torres,
  Nucl.\ Phys.\  B {\bf 782} (2007) 159;
  M.~M.~Nojiri and M.~Takeuchi,
  Phys.\ Rev.\  D {\bf 76} (2007) 015009;
  M.~M.~Nojiri and M.~Takeuchi,
  JHEP {\bf 0810} (2008) 025;
  X.~Wang, Y.~Zhang, H.~Jin and Y.~Xi,
  arXiv:0803.3949 [hep-ph];
  C.~X.~Yue, J.~Y.~Liu, L.~Ding, W.~Liu and W.~Ma,
  arXiv:0811.3267 [hep-ph].
  
\bibitem{Matsumoto:2008fq}
  S.~Matsumoto, T.~Moroi and K.~Tobe,
  Phys.\ Rev.\  D {\bf 78} (2008) 055018.
  
\bibitem{Cao:2007pv}
  Q.~H.~Cao and C.~R.~Chen,
  Phys.\ Rev.\  D {\bf 76} (2007) 075007.

\bibitem{littlest_review}
  M.~Schmaltz and D.~Tucker-Smith,
  Ann.\ Rev.\ Nucl.\ Part.\ Sci.\  {\bf 55} (2005) 229;
  M.~Perelstein,
  Prog.\ Part.\ Nucl.\ Phys.\  {\bf 58} (2007) 247.

\bibitem{littlest}
  G.~Burdman, M.~Perelstein and A.~Pierce,
  Phys.\ Rev.\ Lett.\  {\bf 90} (2003) 241802
  [Erratum-ibid.\  {\bf 92} (2004) 049903];
  T.~Han, H.~E.~Logan, B.~McElrath and L.~T.~Wang,
  Phys.\ Rev.\ D {\bf 67} (2003) 095004;
  M.~Perelstein, M.~E.~Peskin and A.~Pierce,
  Phys.\ Rev.\ D {\bf 69} (2004) 075002.
  
\bibitem{Chen:2006cs}
  C.~R.~Chen, K.~Tobe and C.~P.~Yuan,
  Phys.\ Lett.\  B {\bf 640} (2006) 263.

\bibitem{Komatsu:2008hk}
  E.~Komatsu {\it et al.}  [WMAP Collaboration],
  arXiv:0803.0547 [astro-ph].

\bibitem{LEPSLC}
  ALEPH, DELPHI, L3, OPAL, SLD Collaborations, LEP Electroweak Working Group, 
  SLD Electroweak Group, and SLD Heavy Flavor Group,
  Phys.\ Rept.\  {\bf 427} (2006) 257.

\bibitem{de Austri:2006pe}
  R.~R.~de Austri, R.~Trotta and L.~Roszkowski,
  JHEP {\bf 0605} (2006) 002.

\bibitem{Hubisz:2005tx}
  J.~Hubisz, P.~Meade, A.~Noble and M.~Perelstein,
  JHEP {\bf 0601} (2006) 135.

\bibitem{Arguin:2005cc}
  J.~F.~Arguin {\it et al.}  [CDF Collaboration],
  arXiv:hep-ex/0507091.

\bibitem{Yao:2006px}
  W.~M.~Yao {\it et al.}  [Particle Data Group],
  J.\ Phys.\ G {\bf 33} (2006) 1.

\bibitem{Jungman:1995df}
  For reviews,\\
  G.~Jungman, M.~Kamionkowski and K.~Griest,
  Phys.\ Rept.\  {\bf 267} (1996) 195;
  L.~Bergstrom,
  Rept.\ Prog.\ Phys.\  {\bf 63}, (2000) 793;
  G.~Bertone, D.~Hooper and J.~Silk,
  Phys.\ Rept.\  {\bf 405} (2005) 279;
  C.~Munoz,
  Int.\ J.\ Mod.\ Phys.\ A {\bf 19} (2004) 3093.
  
\bibitem{madgraph} http://madgraph.hep.uiuc.edu/.
\bibitem{physsim} http://acfahep.kek.jp/subg/sim/softs.html.
\bibitem{helas} H. Murayama, I. Watanabe, K. Hagiwara, KEK-91-11, (1992) 184.
\bibitem{bases} T. Ishikawa, T. Kaneko, K. Kato, S. Kawabata,\emph{Comp, Phys. Comm.} {\bf 41} (1986) 127.
\bibitem{pythia} T. Sj$\dot{\mathrm{o}}$strand, \emph{Comp, Phys. Comm.} {\bf 82} (1994) 74.
\bibitem{tauola} http://wasm.home.cern.ch/wasm/goodies.html.
\bibitem{glddod} GLD Detector Outline Document, arXiv:physics/0607154.

\bibitem{Kolb:1990vq}
E.~W.~Kolb and M.~S.~Turner,
{\it The Early Universe},
(Addison-Wesley, Reading, MA, 1990).

\bibitem{GarciaAbia:1999kv}
  P.~Garcia-Abia and W.~Lohmann,
  Eur.\ Phys.\ J.\ direct C {\bf 2} (2000) 2;
  N.~T.~Meyer and K.~Desch,
  Eur.\ Phys.\ J.\  C {\bf 35} (2004) 171;
  P.~Garcia-Abia, W.~Lohmann and A.~Raspereza,
  arXiv:hep-ex/0505096;
  F.~Richard and P.~Bambade,
  arXiv:hep-ph/0703173.

\bibitem{Planck:2006uk}
    [Planck Collaboration],
  arXiv:astro-ph/0604069.
  
\bibitem {Ref:subgroup}
http://www-jlc.kek.jp/subg/physics/ilcphys/.

\end{thebibliography}
